%% file: offlArXiv.tex
\documentclass{article}
\pdfoutput=1

\usepackage[T1]{fontenc}
\usepackage{amsmath,amssymb}

\usepackage{caption} 
\usepackage{subcaption}
\usepackage{graphicx}
\usepackage[colorlinks=true]{hyperref}
\usepackage[utf8]{inputenc}

\newcommand{\schema}{{\textsc{OFFl}}} 
\newcommand{\software}{\textsc{OFFlMaker}}
\newcommand{\website}{\url{http://modeling.mit.edu}}

\begin{document}

\title{\schema\ models: novel schema for dynamical modeling of biological systems}

\author{C.\ Brandon Ogbunugafor\footnote{Department of Organismic and Evolutionary Biology, Harvard University, Cambridge, MA, USA} \and Sean P.\ Robinson\footnote{Department of Physics, Massachusetts Institute of Technology, Cambridge, MA, USA
} \footnote{Corresponding author: \href{mailto:spatrick@mit.edu}{spatrick@mit.edu}}}
\date{2016 June 7}

\maketitle

\begin{abstract}
Flow diagrams are a common tool used to help build and interpret models of dynamical systems, often in biological contexts such as consumer-resource models and similar compartmental models.  Typically, their usage is intuitive and informal. Here, we present a formalized version of flow diagrams as a kind of weighted directed graph which follow a strict grammar, which translate into a system of ordinary differential equations (ODEs) by a single unambiguous rule, and which have an equivalent representation as a relational database. (We abbreviate this schema of ``ODEs and formalized flow diagrams'' as \schema.) Drawing a diagram within this strict grammar encourages a mental discipline on the part of the modeler in which all dynamical processes of a system are thought of as interactions between dynamical species that draw parcels from one or more source species and deposit them into target species according to a set of transformation rules. From these rules, the net rate of change for each species can be derived. The modeling schema can therefore be understood as both an epistemic and practical heuristic for modeling, serving both as an organizational framework for the model building process and as a mechanism for deriving ODEs. All steps of the schema beyond the initial scientific (intuitive, creative) abstraction of natural observations into model variables are algorithmic and easily carried out by a computer, thus enabling the future development of a dedicated software implementation. Such tools would empower the modeler to consider significantly more complex models than practical limitations might have otherwise proscribed, since the modeling framework itself manages that complexity on the modeler's behalf. In this report, we describe the chief motivations for \schema, carefully outline its implementation, and utilize a range of classic examples from ecology and epidemiology to showcase its features. 
\end{abstract}


\section{Introduction}\label{sec:intro} 


When faced with the collected observations of a natural system, one of the principal tasks for a scientist is to bring some level of understanding or order to the observations, typically by collating the raw facts according to some conceptual framework, a theory.
The framework could be an existing paradigm for describing broad categories of observations with a minimum of principles, or a novel model meant only to capture the quantitative details of a single study, or somewhere in between \cite{kuhn}.
The broad approach has proven especially challenging in biological and medical contexts, where the complexity of most systems precludes a reductionist interpretation of observations in terms of a few fundamental principles. In these cases, the ability to quickly generate and evaluate the space of possible models for a system is of great practical importance for advancing the scientific understanding of the system and applying that understanding to real world situations.


Unfortunately, there seems to exist a correlation between the scientific fields where complexity renders facile model building the most important and those fields whose professional culture is associated with an aversion to the most powerful language in which to express models, that is mathematics  \cite{pachter}.   Conversations about the relative cultures of mathematics and other fields --- in particular biology --- have emerged in several forums in recent years.  Some of them have arisen as debates in major journals,  others as informal discussions taking place mostly in cyberspace \cite{PNAS1,PNAS2}.
Part of this cultural divide involves the question of extent: while all parties might agree that mathematical tools allow useful description of systems and phenomenon, on what problems should such approaches be utilized?


The divide notwithstanding, practitioners in biological fields exhibit an impressive command of the vast natural histories (observations) of the systems they study, and the organizational schema 
commonly used to understand them. If there were clearer, more universal methods to describe biological processes, then cross-disciplinary translation would be easier.  This implies to us that a set of tools which put the power of mathematical modeling into the hands of biological and medical practitioners --- and their students --- which do not require advanced mathematical training would be of great practical importance in helping to bridge the cultural divide. 


Here, we describe an organizational schema which enables one to study and understand dynamical systems of the sort that commonly arise in problems of epidemiology, population growth in both human and ecological (\textit{e.g.}, predator-prey and consumer-resource) systems, chemical kinetics, and the like. In all of these situations, parcels of various dynamical species are transformed into parcels of other species at rates determined by a set of interaction rules that depend on the population values of the species. The dynamics of such systems are well-described by rate equations which are systems of coupled (possibly nonlinearly) ordinary differential equations (ODEs). These ODE models are often referred to as deterministic compartmental models. A representative example of this general class of systems are so-called chemical reaction networks \cite[and references therein]{feinberg}, a term which emphasizes the underlying network character of the system. Analogously, we will refer to the more general class of systems considered here as \emph{interaction networks}, but we will not attempt to define that term with any specific rigor beyond the above description. 

The core of the schema is an organizational framework which enforces a certain mental discipline for the modeler, but which is not inherently mathematical, making it readily accessible to anyone wishing to translate practical knowledge of biological processes into the schema. The remainder of the schema is a set of sophisticated but algorithmic rules --- easily automated --- which translate the model into ODEs solvable by standard techniques. We emphasize that any practical implementation of the schema as a software tool could and should bury these automated rules behind a non-mathematical user interface, separating the scientific modeling process from the mathematics used to represent and solve the models. In contrast, this manuscript is addressed to those researchers who might implement such software tools or otherwise investigate the mathematical structures which underlie the schema, and who therefore require more technical detail than the would-be tool user.

To illustrate the use of this modeling approach, we will first introduce the schema, then explain its theoretical underpinnings, and finally demonstrate its utility with a few classical examples from ecology and epidemiology. In the course of this discussion we hope to make clear that the schema and its automatability empowers the modeler to consider significantly more complex models than practical limitations might have otherwise prescribed. Likewise, tools built with this schema would appear to be ideally suited for teaching dynamical modeling to the biological and medical student communities, where quantitative modeling is currently underserved \cite{MCAT} due to a perceived deficit in prerequisite mathematical training. Nevertheless, we do include here a technical discussion of the mathematical back end of the schema.

\section{Methods}\label{sec:methods}
\subsection{Flow diagrams, ODEs, and the modeling process}\label{sec:flodemodel}


When considering systems which can be understood as interaction networks (as defined above), we would like to have a model development process that lets the modeler follow a procedure similar to the following:
\begin{enumerate}
\item Identify all the dynamical quantities (the different species or compartments) in the system.
\item Identify all the processes (interactions between species, be they biological, ecological, physical, or otherwise) in the system.
\item For each process, identify which quantities are ``consumed'' by the interaction and which quantities they are transformed into.
\item From this accounting of species and interactions (which constitutes the model of the system), move quickly to a set of ODEs that can be analyzed by standard means, either analytically or numerically.
\end{enumerate}
At an abstract level, the above steps describe a large range of specific modeling techniques already in common use across a variety of fields.

\begin{figure}
\centering
\begin{subfigure}[b]{0.45\linewidth}\centering
\includegraphics{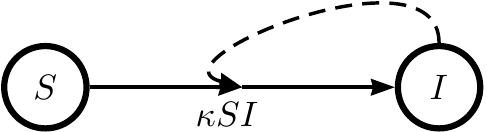}
\caption{}
\label{fig:compMA-OnD}
\end{subfigure}
\begin{subfigure}[b]{0.45\linewidth}\centering
\includegraphics{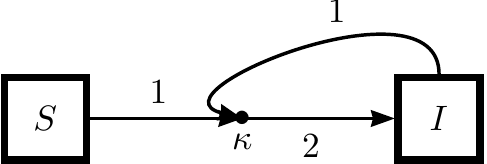}
\caption{}
\label{fig:compMA-new}
\end{subfigure}
\begin{subfigure}[b]{0.45\linewidth}\centering
\includegraphics{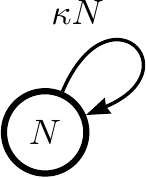}
\caption{}
\label{fig:exp-OnD}
\end{subfigure}
\begin{subfigure}[b]{0.45\linewidth}\centering
\includegraphics{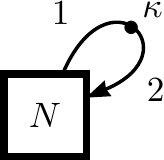}
\caption{}
\label{fig:exp-new}
\end{subfigure}

\caption{{Examples of flow diagrams from two different diagrammatic schema.} Diagrams \subref{fig:compMA-OnD} and \subref{fig:compMA-new} both represent a simple consumer-resource interaction between a ``susceptible'' species $S$ and an ``infectious'' species $I$ --- such as is used in the simplest SI (susceptible-infected) epidemiological models --- but do so in different diagrammatic schema. The edge labels in diagram \subref{fig:compMA-new} further specify a model in which one parcel each of $S$ and $I$ combine to form two parcels of $I$. Both diagrams resolve to the ODE system $dI/dt=-dS/dt=\kappa SI$. Diagrams  \subref{fig:exp-OnD} and \subref{fig:exp-new} both represent exponential growth for a population of size $N$, but do so in different diagrammatic schema.  The edge labels in diagram \subref{fig:exp-new} further specify a model in which each parcel of $N$ splits to form two parcels of $N$. Both diagrams resolve to the ODE $dN/dt=\kappa N$. Diagrams  \subref{fig:compMA-OnD} and  \subref{fig:exp-OnD} are depicted with the widely used schema of \protect{\cite{OnD}}, conceptually representing different kinds of positive feedback processes, while diagrams \subref{fig:compMA-new} and  \subref{fig:exp-new} use the schema described in this document (``\schema'', defined below) and formally represent the routing of parcels of the population through interaction processes.}
\label{fig:compMA}
\end{figure}

 One such technique is the use of ``flow diagrams'' or ``box diagrams'', shown by example in Fig.~\ref{fig:compMA}. As described in a variety of sources \cite[or similar textbooks]{OnD}, flow diagrams have been of enormous practical use in epidemiology and ecology, but their utility is more general than these fields.  
In this technique, each species is written as a small box with the name (or variable name) of the species in it, and directed arrows show the flow of the interactions from one species to another. Individual modelers may apply the technique differently from problem to problem, using it as an informal method for keeping one's thoughts organized. 

One example of a systematized implementation of flow diagrams is that given by the textbook of Otto and Day \cite[in particular  Boxes 2.3, 2.4, and 2.5]{OnD}. In this implementation, the arrows are labeled with the functional form of the interaction rate. In passing from the diagram to ODEs, different interpretive rules are used for arrows which connect species versus arrows that loop back to their source, as in Fig.~\ref{fig:exp-OnD}. Also, different symbology must be used to indicate arrows that are part of qualitatively different interactions. For example, solid lines flowing from the source to the target in a consumer-resource interaction are supplemented by an additional dashed line looping from the target to the midpoint of the the solid arrow, as illustrated in Fig.~\ref{fig:compMA-OnD}.

One might hope that such diagrams could be interpreted loosely as networks or graphs, in the mathematical sense, such that the understanding and utilization of models described by these diagrams could benefit from the tremendous advances that have been made in network theory over the past two decades, as well as from the simultaneous proliferation of software tools now available to analyze these systems. With modern network tools, analysis of truly enormous systems could be possible \cite{barabasi, watts}. Indeed, to describe very complex systems of ODEs such as gene regulation networks or ``all world'' economy-ecology models \cite[for example]{world3}, it is imperative that any flow diagram representation be brought under strict control, if only to avoid errors in the modeling process. However, existing rule schema for constructing and interpreting these diagrams as a system of ODEs often exhibit some level of interpretational ambiguity which must be solved by human intervention, and are not as yet sufficiently regular to bring flow diagrams fully under the purview of network theory.
The new organizational schema for managing ODEs and formalized flow diagrams (``\schema'') presented in the following sections aims to alleviate this difficulty.

\subsection[the section motivating \schema]{Formalizing flow diagrams in the \schema\ schema}

To enable this level of rigor, the \schema\ schema prescribes a formalized version of the modeling process described above, including the representation of interaction network systems as formalized flow diagrams.
In \schema, such systems are thought of as a directed network of species and interactions, with any given interaction having some species as sources and others as targets. (The nodes of the network represent the species and interactions while the edges represent flows of parcels from species to interactions and vice versa. This way of thinking is already a significant departure from common practice with flow diagrams. For example, it forbids edges which enter or leave the network from ``nowhere''.) We therefore want the schema to force the modeler --- motivated by a physical, biological, or ecological understanding of the system being modeled --- to work directly at the level of the dynamical quantities under examination and the processes which affect them, rather than, say, skipping ahead to a particular differential equation or dynamical solution. 
Thus, the structure of the flow diagrams should reflect this low-level approach in a way that can be uniformly applied to all possible processes.

In the diagrammatic schema to be presented here, an arrow leaving a species will represent a negative contribution to that species's rate of change while an incoming arrow will represent a positive contribution. Thus, for example, without any further adornments, an arrow that loops back into its source represents a parcel of the species leaving and then reentering the population, for no net change. This is another departure from typical usage, in which a self-looping arrow represents a  positive feedback, such as in the exponential growth model shown in Fig.~\ref{fig:exp-OnD}. We view this departure from convention as an advantage because even the case of simple exponential growth is not simply a feedback, but rather a process in which each parcel is amplified at each time step by a reproduction process. We would like the diagram to not just to indicate a feedback in the numerical value, but also to capture the physical cause of that feedback. So, in this case, the self-looping arrow should be interrupted by a new node: a ``dot'' representing the reproduction process, as in Fig.~\ref{fig:exp-new}. The dot itself carries the rate constant of the process (literally a constant in the case of exponential growth), while its incoming and outgoing edges carry additional labels showing the proportions of how much incoming ``stuff'' turns into how much outgoing ``stuff''. 

In fact, to achieve a uniform application of the schema in a way that enforces thinking at a process level, \emph{all} interactions between species in the network will be represented by a labeled dot connected to species by arrows. Furthermore, species will \emph{never} be connected to each other except via an interaction. For example, in the simplest SI (susceptible-infected) epidemiological model (illustrated in Fig.~\ref{fig:compMA}), one might informally think of the basic mechanism as following the schematic equation  $S+I \rightarrow I$, but being more careful, we recognize that reality is better reflected by $S+I \rightarrow I+I=2I$. (A critical insight into such systems is that the total number of people is conserved in every encounter!) The former way of thinking about the system is captured in the flow diagram of Fig.~\ref{fig:compMA-OnD}, while the latter way of thinking, which we believe to be more disciplined and rigorous, is expressed in Fig.~\ref{fig:compMA-new}. 

Therefore, in \schema, the SI model shows the infection process as a dot --- labeled with a rate constant --- which draws in parcels from two source species and then outputs parcels to a target species with twice the ``weight''. In contrast, Fig.~\ref{fig:compMA-OnD} simply shows one species flowing directly to the other, representing the infection process as an arrow. The additional dashed line in Fig.~\ref{fig:compMA-OnD} is cosmetic, indicating to the reader that the rate of the process is modulated by another species. This point of departure from typical usage of flow diagrams is worth repeating: in an \schema\ diagram, both species and processes will be represented as nodes of the network while an edge represents a relationship between a species and an interaction, whereas in typical diagrammatic schema, only species appear as nodes while edges represent processes.

The labels on the edges and on the interaction process together contribute to a rule which governs the rate of change for each species involved in the process: how much to take away from each source species and how much to redistribute to each of the targets.  The sum of these contributions to each species over all processes constitutes the dynamics of the system, expressed as a set of ODEs. How one moves from a descriptive understanding of the system to a formal representation as a flow diagram and then to ODEs is the subject of the next several sections of this manuscript.

It is worth noting that rates of interaction --- if left to chance --- depend on the rate of uncorrelated random encounters between parcels of the source species. Therefore, a given interaction's contribution to the system dynamics will be proportional to the value of each of its source species (not the target species), as well as having a contribution inherent to the interaction itself. For example, in SI infection models, the rate of new infection is proportional to $SI$ and additionally proportional to a rate constant describing how often $S$ and $I$ encounters which lead to new infection occur. This proportionality is part of the kinematic structure of such models --- a mass action law arising from an underlying dynamics in the uncorrelated microscopic degrees of freedom, consisting of a random walk or diffusion with localized ``billiard ball'' interactions. Deviations from this proportionality reflect interesting dynamics of the interaction process beyond that of simple randomly walking parcels. (For example, transmission rates of sexually transmitted diseases or predation rates in ecological systems with ample prey are not simply proportional to the values of the involved populations because the interaction event itself can only occur a maximum number of times in any fixed time period.) Given this, and given that we want the flow diagram to distinguish the modeling of the interaction process from the basic kinematic structure of population dynamics, the label on an interaction process in an \schema\ diagram should specify the fractional rate of transformation of the source species (\textit{e.g.}, per capita change per hour) such that the absolute transformation rate (\textit{e.g.}, quantity per hour) which contributes to the final ODEs will be given by the product of this fractional rate with the population values of each source species.

\subsection{\schema\ schema: from a system model to a diagram}\label{sec:diagram}
Based on the above discussion of general principles, we now propose a specific process for developing a model and representing it as a flow diagram as follows. (Terms introduced as jargon of the schema are indicated in \emph{italic} typestyle. They are collected and defined in Appendix~\ref{sec:defs}.)
\begin{enumerate}
\item Think carefully about the system, about which features are important for modeling and which can be ignored, and about how the important features can be quantified as measurable numbers. Decide which aspects of the system are dynamical quantities --- that is, changing in time in response to the values of the other dynamical quantities --- and which aspects are the processes that cause those dynamical changes. (The dynamical quantities of the system will be represented by the species of the model. The processes of the system will be represented by the interactions of the model.)
\item Consider each process to be drawing \emph{parcels} from certain dynamical quantities at some rate, transforming them, and then depositing those transformed parcels back into other dynamical quantities. Any process being regarded as an external source or sink for parcels should instead be included as a dynamical quantity of the system.
\item Make a list of the \emph{species} (the dynamical quantities) involved in the model. 
\begin{enumerate}
\item Assign a variable name to each species.
\item Draw a box with the variable name in it for each species.
\end{enumerate}
\item Make a list of the \emph{interactions} (the different processes that define how the values of the species change in time) involved in the model. 
\begin{enumerate}
\item Assign variable names to the properties associated with each interaction.
\begin{itemize}
\item Identify which species are \emph{sources} for the interaction (having parcels taken away by the interaction at each time step) and which are \emph{targets} (receiving parcels).
\item Determine what size of parcel is drawn from each source species by each application of the interaction. These numbers are called the \emph{source weights}.
\item Determine what size of parcel size is delivered to each target species by each application of the interaction. These numbers are called the \emph{target weights}.
\item Determine what this interaction contributes to the fractional rate of change for each parcel of source species (\textit{e.g.}, per capita change per hour, or the time derivative of the logarithm). In particular, how does the fractional rate of change scale with the values of each species in the system? (A lack of scaling --- a numerical constant --- is common in many models.) This functional dependence of the rate on the species is called the \emph{interaction function}. It is based on the modeler's empirical knowledge of the process and the system.
\end{itemize}
\item  Draw a point for each interaction and label it with the interaction function.
\item Connect each interaction to its target and sources species.
\begin{itemize}
\item  Draw an arrow from each source species box to the interaction dot. Label it with that species's source weight unless the weight is 1.
\item Draw an arrow from the interaction dot to each target species box. Label it with that species's target weight unless the weight is 1.
\end{itemize}
\end{enumerate}
\end{enumerate}

Summarizing, each species is represented by a box with the variable name inside.  The interaction is represented by a small dot labeled by the interaction strength (or interaction function, if it is not a constant). All source species for the interaction get an arrow from the box to the dot. All target species get an arrow from the dot to the box. If some of the source species contribute more than others, then the source arrows may be labeled with a weighting factor. If the targets do not receive equal fractions, or if the targets receive a different amount than is output from the sources, then the target arrows may be labeled with a weighting factor. Any unlabeled arrows are assumed to have weight 1.  

The weights and interaction functions could be functions of time, or of the dynamical species values, or of external parameters, but in practice they will often be constants. However, the interaction function in particular will sometimes reflect a nontrivial functional dependence on the species values, for example being inversely proportional to the sum of all the species values. Weights may be pure numbers or they may have units, for example, ``hares consumed per lynx birth'' in a predator-prey system. Also, there is no requirement that weights into and out of an interaction be in balance: an interaction may represent a net gain of parcels, for example.
Further discussion of the interpretation of weights and interaction functions is given in Appendix~\ref{sec:weightsNstuff}.
 
Ultimately, a model with $M$ species and $N$ interactions will result in a network with $M+N$ nodes: $M$ boxes and $N$ dots. Each box connects to some dots by a weighted directed edge outgoing to the dot for source boxes and incoming to the box for target boxes. Boxes cannot connect to boxes. Dots cannot connect to dots. Boxes are labeled with the species name. Dots are labeled with the interaction rate. All dots must have at least one incoming and one outgoing edge. All edges must terminate at one end on a box and at the other on a dot. ``Loose ends'' are not allowed. ``Direct feeds'' from one box to another --- that is, transformations of a species without causation from an interaction --- are not allowed. ``Splits and merges'' which connect dots without an intermediate box are not allowed.

\subsection{\schema\ schema: from a diagram to ordinary differential equations}\label{sec:equations}
The interpretation of a given diagram as an ODE proceeds by the following rules:
\begin{itemize}
\item A network with $M+N$ nodes in which $M$ nodes are labeled as dynamical variables $X^i(t)$ ($i\in\{0\ldots M-1\}$) and the remaining $N$ nodes are labeled with interaction functions $f_a$ ($a\in\{0\ldots N-1\}$) will represent a system of $M$ first order ODEs, \textit{i.e.}\ a system of the form $dX^i/dt = F^i(X)$ for some set of functions $F^i$ to be determined as described below.
\item The species nodes and interaction nodes are connected by directed labeled edges. For each interaction $f_a$, note the number $p_a$ of outgoing \emph{target edges} and the number $q_a$ of incoming \emph{source edges}. 
\item Construct the array $\alpha_{a,m}$ which is the weight on the $m$\textsuperscript{th} target edge of the $a$\textsuperscript{th} interaction and the array  $\beta_{a,n}$ which is the weight on the $n$\textsuperscript{th} source edge of the $a$\textsuperscript{th} interaction.
\item Also construct the \emph{target array}
\begin{equation}
t^i_{b,m}=\left\{\begin{array}{cl}1 &\textrm{if $X^i$ is the target of the $m$\textsuperscript{th} edge of interaction $a$ } \\ 0 &\textrm{otherwise}\end{array} \right.
\end{equation}
and \emph{source array}
\begin{equation}
s^i_{b,m}=\left\{\begin{array}{cl}1 &\textrm{if $X^i$ is the source of the $m$\textsuperscript{th} edge of interaction $a$ } \\ 0 &\textrm{otherwise}\end{array} \right. .
\end{equation}
\item Then, 
\begin{equation}
\frac{dX^i}{dt}=\sum_{a=0}^{N-1}{\left[
          \left(  \sum_{m=0}^{p_a-1}{\alpha_{a,m}t_{a,m}^{i}} 
                - \sum_{n=0}^{q_a-1}{ \beta_{a,n}s_{a,n}^{i}} \right) f_a
          \prod^{q_a-1}_{\ell=0}{\left(
                              \sum_{j=0}^{M-1}{ s_{a,\ell}^{j}X^{j} }
                              \right)}
     \right]}. \label{eq:central}
\end{equation}
\end{itemize}
This is the central result: an ODE which can be solved by standard techniques. Note that care should be taken not to confuse the time variable $t$ and the target array $t^i_{b,m}$.


It is interesting to note that different diagrams can result in the same ODE, but the model interpretation of those diagrams as ``systems of processes that redistribute transformed parcels among species'' could be different. In the language of mathematical physics, 
transformations of model quantities (such as elements of the diagram) which leave the behavior of the model (that is, Eq.~\ref{eq:central}) unchanged are known as ``symmetries'' of the system.  Symmetries generally indicate a redundancy of description in the model and carry interesting qualitative consequences for the system's dynamics, such as the existence of conserved quantities \cite{noether, AJP1, AJP2}. For example, in electrical circuit networks, the modeler's freedom to choose the ground point at which voltage is zero is related in a deep way to the conservation of electric charge, which in turn leads to Kirchhoff's laws for electrical currents: powerful relationships which constrain the possible behaviors of quantities in the network \cite{kirchhoff}. Though intriguing, for the sake of clarity we will not further explore these concepts in the current manuscript.

\subsection{Organizing \schema: From a diagram to a database representation}
The various quantities described in the previous section which specify the diagram also define a relational database consisting of four tables: one each for the species nodes, the interaction nodes, the source edges, and the target edges. We prove this claim by explicit construction of the tables as shown in Fig.~\ref{tab:DBtab}.
\begin{figure}
{\footnotesize \sffamily
\begin{center}
\begin{subtable}{.5\linewidth}
\input{node-table.tex}
\end{subtable}
\\  \smallskip

\begin{subtable}{.5\linewidth}
\input{interaction-table.tex}
\end{subtable}
\\ \smallskip

\begin{subtable}{.5\linewidth}
\input{target-table.tex}
\end{subtable}
\\ \smallskip

\begin{subtable}{.5\linewidth}
\input{source-table.tex}
\end{subtable}
\end{center}
\caption{The four tables which represent \schema\ diagrams as a relational database. The ``Example'' columns refer to quantities as they appear in Eq.~(\ref{eq:central}).}\label{tab:DBtab}

}
\end{figure}
 In the language of relational databases, the source table and target table act as linking tables between the species table and interaction table. 

Each table could of course be decorated with additional fields, like textual descriptions of the information being represented, or fields which support technical execution within a particular software implementation. For example, a \textsc{key} field consisting of a concatenation of the \textsc{interaction-index} and \textsc{edge-index} fields could be added to the target and source tables. Or, a new table uniquely keyed by \textsc{node-index} and the time variable could be used to keep track of the changing value of \textsc{species-value}, instead of dynamically updating the field in the node table.

The elements of an \schema\ flow diagram and its database representation are in exact one-to-one correspondence, so the procedure given above for generating a diagram also applies to generating the database, and either the diagram or the database can be used equivalently in generating Eq.~(\ref{eq:central}).  The database representation is readily stored and interpreted by a computer, making the generation and subsequent solution of Eq.~(\ref{eq:central}) for a given system a directly automatable process. 

\subsection{Examples of simple processes as \schema\ models}\label{sec:simple}
Many examples of practical importance involve only the linear processes of exponential growth and death along with bilinear (that is, nonlinear through the product of two variables) mass action processes. As the core building blocks of many complex models, we consider these three processes here in some detail.

\begin{figure}
\begin{subfigure}[b]{0.45\linewidth}\centering
\includegraphics{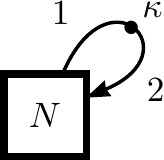}
\caption{}
\label{fig:exp-gro}
\end{subfigure}
\begin{subfigure}[b]{0.45\linewidth}\centering
\includegraphics{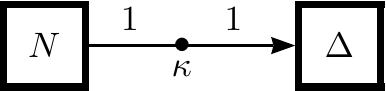}
\caption{}
\label{fig:exp-det}
\end{subfigure}
\caption{{Exponential growth and death in the \schema\ schema.} \subref{fig:exp-gro} Growth is shown as positive feedback process. \subref{fig:exp-det} Death is shown as a change of state.}
\label{fig:exps}
\end{figure}

Despite the processes of exponential growth (as shown in  Fig.~\ref{fig:exp-new}) and death being represented with the same ODE, differing only by the sign of the coefficient, the biological processes which give rise to these phenomena are quite different. Appropriately, they are represented quite differently as flow diagrams. In both processes, a parcel is drawn from the population $N$ at rate $\kappa$. In growth, that parcel is replicated (say, duplicated, as shown in Fig.~\ref{fig:exp-gro}) and returned to the population. In death, however, the parcel is simply moved from a live state population to a dead state population $\Delta$, as shown in Fig.~\ref{fig:exp-det}. Thus, growth is given as $dN/dt = k(-1+2)N = \kappa N$, while death is given by the system $dN/dt= -\kappa N$ and $d\Delta/dt=+\kappa N$. 

The formal grammar of \schema\ diagrams does not allow an edge to end without a target node. Therefore, the existence in Fig.~\ref{fig:exp-det} of the target species $\Delta$ and its dynamical equation is required, but the dynamical equation for $N$ remains independent of $\Delta$ and can be solved without the solution for $\Delta$. The existence of such auxiliary variables --- representing an external reservoir whose dynamics are unimportant to the system variables --- is typical for \schema\ models of ``open'' systems that exhibit processes similar to death, immigration, or emigration: processes that might be represented in a less formal schema by lines that enter or leave the diagram from nowhere. On the other hand, a process like death could perhaps be modeled instead as a spontaneous eradication rather than as a change of state. In this case, the diagram for the death process would take same form as for the growth process as in Fig.~\ref{fig:exp-gro}, except with a 0 instead of a 2 on the target edge, resulting in the single equation $dN/dt=-\kappa N$ and no auxiliary variable $\Delta$ representing a death state.

\begin{figure}
\centering
\includegraphics{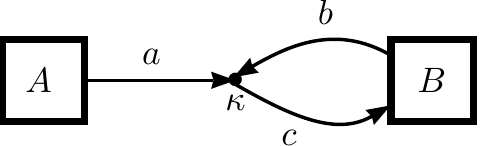}
\caption{{ \schema\ diagram for the most general two-species consumer-resource process.} }
\label{fig:genMA}
\end{figure}

The two-species consumer-resource interaction is perhaps the simplest nonlinear process appearing in the types of models under consideration here. One notable example, the SI epidemiological model, has already been presented in Fig.~\ref{fig:compMA-new}. The most general version of the process, in which $a$ parcels of species $A$ combine with $b$ parcels of species $B$ at rate $\kappa$ to become $c$ parcels of species $B$  is shown in Fig.~\ref{fig:genMA}. Its associated ODE is 
\begin{subequations}
\begin{align}
dA/dt &= -a\kappa AB\\
 dB/dt &= (c-b)\kappa AB .
\end{align}
\end{subequations}
Note, if the number of parcels incoming to the interaction equals those outgoing ($a+b=c$) as in the SI system, then $\frac{d}{dt}(A+B)=0$, and the total population is therefore constant. Otherwise, the consumer-resource process will lead to overall growth or death of the population.

\section{Discussion}\label{sec:example}

In this section, we will discuss two slightly more complex examples. Both 
are established models whose usage is well understood in several scientific communities. This provides an opportunity to demonstrate how \schema\ works and affirms that \schema\ can recapitulate results from well characterized biological problems. Additional examples will be given at \website.

\subsection{Susceptible-Infected-Recovered (SIR) system}\label{sec:SIR}
In an epidemiological SIR system \cite{kermack}, susceptibles $S$ contact the infected $I$ at a rate $c$ (say, encounters per infected per day) of which a fraction $a$ of contacts lead to new infections, while the infected spontaneously recover at a rate $\rho$ (say, recoveries per infected per day), becoming the resistant population $R$. The specific system considered here also includes the waning of immunity at rate $\sigma$ (losses per resistant per day), returning the resistant back to the susceptible population; death rates $d$ and $\delta$ (deaths per person per day) for each species; and immigration of new susceptibles at rate $\theta$ (susceptibles per day).

\begin{figure}
\centering
\begin{subfigure}[b]{\linewidth}\centering
\includegraphics{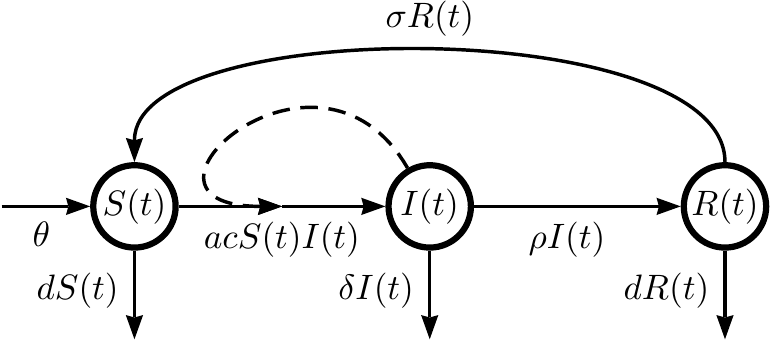}
\caption{}
\label{fig:SIR-OnD}
\end{subfigure}

\begin{subfigure}[b]{\linewidth}\centering
\includegraphics{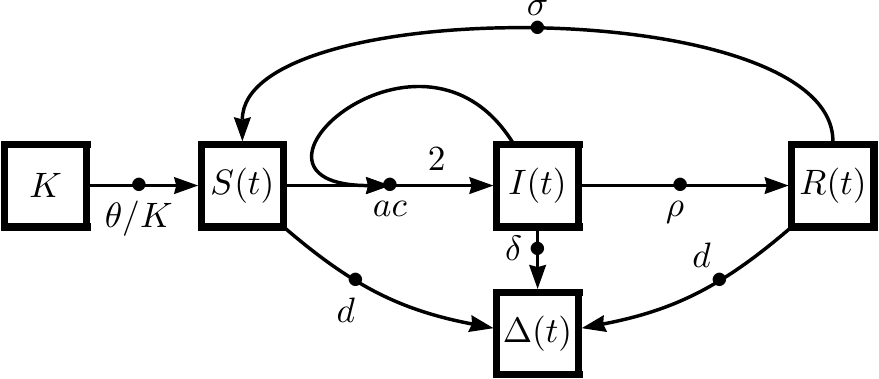}\phantom{\hspace{.75 in}}
\caption{}
\label{fig:SIR-new}
\end{subfigure}

\caption{{A simple epidemiological SIR model.} Contact between susceptibles $S$ and infected $I$ leads to new infections, while simultaneously the infected can spontaneously recover and become resistant to the infection $R$. \subref{fig:SIR-OnD} An informal flow diagram for the SIR model as shown, for example, in Fig.~3-9 of Ref.~\cite{OnD}. \subref{fig:SIR-new} The \schema\ version of SIR. The new ``dead pool'' species $\Delta$ is the target of the death interactions. The new external ``community pool'' species $K$ is the source of the immigration interaction. The model has 5 species and 7 interactions. }
\label{fig:SIR}
\end{figure}

The SIR system is ideally suited for modeling as a flow diagram. Fig.~\ref{fig:SIR} shows a side-by-side comparison of two different schema representations of this SIR model: a traditionally informal flow diagram and an \schema\ diagram. The \schema\ diagram is generated by starting from the above description of the system and following the steps given in Section~\ref{sec:diagram}.

Given the diagram, the rules leading to Eq.~(\ref{eq:central}) then illustrate how to generate a set of ODEs, which proceeds as follows.
We first identify all the species and name them: along with the three dynamic species $S$, $I$, and $R$, we also add a ``community pool'' species $K$ which is the source for external immigration and a ``dead pool'' species $\Delta$ which is the target for death interactions. 
Next, reasoning through the description of the model given above, we see there are seven interactions in the model: the death of each of the three dynamical species (moving a unit parcel from $S$, $I$, or $R$ to $\Delta$ with rate $\delta$ or $d$), spontaneous recovery (moving a unit parcel from $I$ to $R$ with rate $\rho$), loss of immunity (moving a unit parcel from $R$ to $S$ with rate $\sigma$), infection (in which a parcel of $S$ interacts with a parcel $I$ at rate $ac$ to become two parcels of $I$), and immigration (moving a unit parcels from $K$ to $S$). In immigration, $\theta$ --- the absolute rate of change to $S$ --- is presumed by the model to be constant even as the number of available parcels in the $K$ population varies, so it must the case that the strength of the process which attracts parcels from $K$ to $S$ must vary as $\sim 1/K$, specifically $\theta/K$. The dynamics of $S$, $I$, and $R$ are then independent of the dynamics of $K$.

The resulting ODE is then
\begin{subequations}
\begin{align}
dS/dt &= -ac SI           +\sigma R   - d S + \theta  \label{eq:SIR-S}\\
dI/dt &=  ac SI  - \rho I             - \delta I  \label{eq:SIR-I} \\
dR/dt &=         + \rho I  -\sigma R  - d R \label{eq:SIR-R}\\
dK/dt &= -\theta \label{eq:SIR-K} \\
d\Delta/dt &= dS+ \delta I + dR . \label{eq:SIR-D}
\end{align}
\end{subequations}
Eqs.~(\ref{eq:SIR-S}), (\ref{eq:SIR-I}), and (\ref{eq:SIR-R}) are the essential dynamics of the system. These are the equations which would result from a standard reading of the informal diagram in Fig.~\ref{fig:SIR-OnD}.
As discussed in Section~\ref{sec:simple}, \schema\ results in two additional equations for $K$ and $\Delta$, but these are trivial additions to the dynamics which can be ignored when solving for  $S$, $I$, and $R$. 

Note that to maintain realism in the model, the initial condition on $K$ should be sufficiently large that $K$ will not change appreciably during the run time $\tau$ of the model, $K(0)\gg\theta\tau$. Or, alternatively, one could force $K$ to be constant  $dK/dt=0$ by any number of ways while leaving the rest of the ODE unchanged, such as setting the immigration source weight to 0.

\subsection{Lotka-Volterra predator-prey system}

\begin{figure}
\centering
\includegraphics{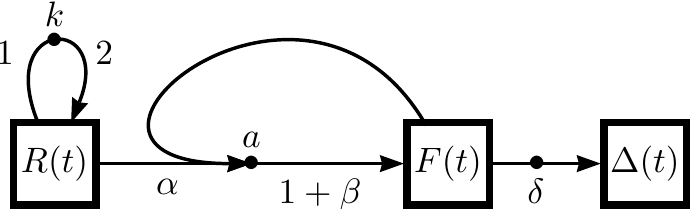}
\caption{{Lotka-Volterra flow diagram.} The Lotka-Volterra model is the simplest predator-prey model. A prey species $R$ (``rabbits'') reproduce spontaneously at rate $k$ while a predator species  $F$ (``foxes'') consume $\alpha$ rabbits at rate $a$ and convert them to $\beta$ new foxes. Foxes move spontaneously at rate $\delta$ to the dead pool $\Delta$.\label{fig:lot-volt}}
\end{figure}

The Lotka-Volterra system is one of the defining models in ecology, and has served as a basis for understanding predator-prey dynamics for many years \cite{lotka, volterraOrig, volterra}. Although superseded by more modern models in ecological research, Lotka-Volterra remains a classic application of mathematical modeling. The \schema\ treatment of Lotka-Volterra is shown in Fig.~\ref{fig:lot-volt}, with the system itself described in the figure caption. Following the steps leading to Eq.~(\ref{eq:central}), Fig.~\ref{fig:lot-volt} renders to the following equations:
\begin{subequations}
\begin{align}
dR/dt &= kR - a\alpha RF \\
dF/dt &= a\beta  RF - \delta F\\
d\Delta/dt &= \delta F . \label{eq:lot-lot}
\end{align}
\end{subequations}
As discussed in Section~\ref{sec:simple}, modeling the death of the predator species $F$ again involves the addition of an auxiliary species $\Delta$ whose dynamics are ultimately irrelevant to the other dynamical species.

In the language of mathematical ecology, the ``functional response'' of the model in this case is so-called type~I. That is, the rate of prey consumption is linear in prey population \cite{holling}, indicating that predators do not eat any less often as they consume prey: they never get ``full''. Changing to a type~II functional response --- which saturates to a constant as the predators get full --- is as simple as changing $a$ in Fig.~\ref{fig:lot-volt} to a function of $F$ and $R$, say $a(F,R)=a/(1+haR)$ for some ``handling time'' $h$. This easy extensibility of the simple model into a more complicated one is typical of the \schema\ schema.

\section{Conclusions}
The goal of this manuscript was to describe a framework which puts the power of mathematical modeling into the hands of content experts who might otherwise avoid the use of highly mathematical tools. We focused on a particular framework that manages systems well-described by ODEs (``interaction networks''), but the general program should be considered open for tools which address other mathematical methods (such as stochastic systems) in a similar spirit.
Specifically, we have attempted the following:
\begin{itemize}
\item Introduce the \schema\ schema, explaining both how it works and its theoretical underpinnings
\item  Highlight differences between \schema\ and related approaches to modeling biological systems which also use flow diagrams and ODEs
\item Illustrate how \schema\  might offer advantages in modeling dynamical systems

\end{itemize}


Mathematically, we have shown that the broad class of dynamical systems under consideration here can be represented in three complementary ways: a weighted directed graph (a network), a system of ODEs, and a relational database. While fundamentally interesting and a clear opportunity for future research into the properties of these systems, further mathematical study of this observed triality is beyond the scope of the present manuscript and will be left for future communications.


Algorithmization of the model building process naturally lends itself to a software engineering implementation. One could imagine a graphical front end that allows a would-be modeler to draw an \schema\ diagram intuitively --- based, for example, on the modeler's observations of infectious disease dynamics in the field ---  while all the mathematics is carried out behind the scenes. This can put the modeling of even complicated outbreaks into the hands of front line medical personnel and public health officials
while requiring little mathematical expertise. Development of such a software implementation of \schema\ as a free web application \cite{modeling} is presently underway and will be the subject of a future report.
Such a software solution could also enable improved instructional methods in and appreciation for the value of quantitative modeling in the biological, public health, and medical student communities.

Because the \schema\ approach lends itself so readily to automation, it is tempting to compare it to the myriad of existing software packages and simulators that model complex systems in biology. Before making any direct comparisons, however, we should reiterate that this manuscript is intended as neither a user manual nor a rationale for a pending software package. Instead, \schema\ is designed and presented as an epistemic framework for understanding how modeling works in biological systems. The true goal of this approach, then, is not to make the act of simulating a dynamic system easier through software, but rather to support the systematic use of systems thinking and the iterative mental processes of model building (and, perhaps, modeling education).

We might even say that \schema\ could be compatible with existing software packages in the systems biology and epidemiology communities.  We suggest that users of these software tools make use of the \schema\ framework in constructing models (perhaps through ``pen and paper'' methods) before using any given package or simulator.  Additionally, for instructors and students, \schema\ is a way of introducing modeling concepts prior to using any software package.

Nevertheless, a modeler might choose an \schema-based software tool for a variety of reasons. Existing software packages for modeling in systems biology and related fields tend to have applications focused on specific classes of problems, such as \textsc{VCell} \cite{vcell} (cell biology, biochemistry), \textsc{CellDesigner} \cite{celldesigner,celldesigner1,celldesigner2} (biochemistry and gene regulation), and \textsc{EpiModel} \cite{epimodel} (modeling of epidemics). In this sense, \schema\ differs in its purpose and fungibility. It can be used for any given systems-style model in ecology, evolution, epidemiology, chemical kinetics, or even social sciences. Because of this, \schema\ might be compared to software packages like \textsc{STELLA} \cite{stella} or \textsc{Berkeley~Madonna} \cite{madonna}, both highly developed and well-regarded systems simulation software packages whose developers also have a mind towards making modeling accessible to novice users and promoting widespread usage of systems thinking.  But while \textsc{STELLA} and \textsc{Berkeley~Madonna} have many potential applications (including in social sciences and business), they are not free packages. Again, it may be possible to implement the \schema\ schema within any of the above software tools, but for an \schema-based software tool to not miss the point of using the \schema\ schema, it should take care to present a user interface which only requires graphical manipulation of the species and interactions of the model, avoiding any user intervention requiring advanced mathematical or computational training.

Several authors have noted the importance of universal standards for the representation and transmission of information about models in systems biology, resulting in successful projects such as the Systems Biology Markup Language (SBML) \cite{SBML} and Systems Biology Graphical Notation (SBGN) \cite{SBGN}. \schema\ can work within these modern standards; in particular, we note that \schema\ diagrams are readily represented using the entity pool nodes, process nodes, consumption arcs, and production arcs of the SBGN Process Description language (SBGN-PD) \cite{SBGN-PD}. (However, modulation arcs from SBGN-PD do not have a direct interpretation under the rules of \schema, which requires that modulations such as catalysis be either explicitly modeled or represented in an interaction function.) However, we must again emphasize that \schema\ differs from strictly symbolic notational frameworks like SBGN because it is more than a visual standard for depicting qualitative biological relationships. Instead, \schema 's elements have dedicated mathematical definitions, and consequently, can be used to model quantitative relationships between actors (biological, social, or other), not simply illustrate them. 

In the end, improving the community's general mathematical literacy is a daunting task that requires commitment and investment from educators, politicians, and active researchers alike. 
That said, even a small pedagogical breakthrough can go a long way --- by being engaged, improved upon, and modified --- towards bridging gaps between complicated mathematics and real world applications.

The \schema\ modeling approach presented here aims to improve our understanding of dynamical systems modeling, a tool that is increasingly useful to both practitioners of science and citizen-scientists in an increasingly complex world.

\section*{Acknowledgements}
The authors wish to thank the participants and organizers of the 2013 Summer School of the New England Complex Systems Institute, which provided some initial inspiration for this project. In particular, we acknowledge Gabriela Michel and Urbano França. SPR also acknowledges the participants of the 2014 Gordon Conference on Physics Research and Education for many thorough and frank conversations on the complex cultural interface of the biological and physical sciences. CBO would like to acknowledge the organizers and participants in the program entitled ``Evolution of Drug Resistance" at the Kavli Institute for Theoretical Physics at the University of California, Santa Barbara (2014) for an immersive experience with regards to multidisciplinary approaches to biomedical problems.
CBO acknowledges the Ford Foundation for funding. SPR acknowledges funding for publishing support from MIT Libraries.

\appendix

\section{Definitions}\label{sec:defs}
Definitions of terms introduced as jargon of the \schema\ schema are as follows. 
\begin{description}
\item[interaction network] The type of system that the \schema\ schema can describe, modeled by a system of coupled ODEs. Parcels of various dynamical species in the model are transformed into parcels of other species at rates determined by a set of interaction rules that depend on the population values of the species.
\item[parcel] The unit of a species which participates in each application of an interaction. Indicated on an \schema\ diagram by the weight labels.
\item[species] The model representation of the dynamical quantities of a system. Species are changed over time in response to the other species by the interactions. Represented in \schema\ by a box labeled with the species name.
\item[interaction] The model representation of the processes of a system which cause the changes in the species. Represented in \schema\ as a dot labeled by the interaction function.
\item[source] Species from which parcels are drawn by an interaction. Represented in \schema\ as a directed edge pointing from a species to an interaction, labeled by a weight.
\item[target] Species to which parcels are delivered by an interaction. Represented in \schema\ as a directed edge pointing from an interaction to a species, labeled by a weight.
\item[weights] Label on an edge of an \schema\ diagram indicating the parcel size for the species drawn out or delivered by the interaction.
\item[interaction function] The rate of change per unit of source species (exclusive of per-species weighting) induced in each species involved in an interaction. Used to label the interaction dots in an \schema\ diagram.
\end{description}
The following terms are not part of the \schema\ schema, but are included here to clarify their meaning as used in this document.
\begin{description}
\item[system] The subject under study which physically exists in the real world (or possibly a hypothetical world) and can be probed by observation and experiment.
\item[model] A representation of a way of understanding the system. May be mathematical, taxonomical, \textit{etc.}  Essentially all modes of understanding systems can be considered models, and essentially all models describe only a fraction of the system's behavior or composition. The value of a model is judged by how faithfully it represents the aspects of the system which are of interest to the modeler. In \schema, the system is represented in three mathematically equivalent ways: a diagram, a database, and a set of ODEs.
\item[process] Mechanism by which a system changes in time. In \schema, processes are represented as interactions which move parcels between species.
\end{description}

\section{Discussion of interaction functions and the meaning of edge weights in an \schema\ model}\label{sec:weightsNstuff}

The interpretation of an \schema\ diagram follows the simple rule that an arrow leaving a box reduces the value of that species while an arrow entering a box increases it. The interaction function determines the interaction's contribution to the fractional rate of change relative to the source terms for all species connected to the interaction. That is, the absolute rate of change due to the interaction is the product of the interaction strength and the values of all the source species. Each species connected to the interaction then gets a contribution to its instantaneous rate of change given by the product of the above absolute rate with the species's weight factor for that interaction, keeping in mind that the contribution to source terms is negative.  So, the total rate of change for a species is a sum of terms, one for each interaction affecting the species.

The source and target weights can be thought of as a kind of conversion factor between species. For example, if a model describes an organism consuming some amount of food before reproducing, the source edge from the food species to the reproduction interaction would have a weight indicating how much food must be consumed to produce each new organism (that is, for each application of the interaction function), while the source and target edges between the reproduction interaction and the organism will be weighted with values 1 and 2, respectively, indicating that the reproduction process results in two organisms for each initial organism participating (binary fission). (The diagram for this model would be similar to the consumer-resource process shown in  Fig.~\ref{fig:genMA}, with $A$ as the food, $B$ as the organism, $a$ as the quantity of food to be consumed, $b=1$, and $c=2$.) In this case, the weight of 2 on the target edge can be viewed as kind of amplification or gain factor, which is a valid alternative point of view for understanding the source and target weights. This illustrates the point that interactions need not be ``conservative'' or ``unity gain'' in \schema: a unit parcel of source species need not be converted into a unit parcel of target species.

\input{offlArXiv.bbl}
\end{document}

%% file: node-table.tex
\begin{tabular}{|l|l|l|}
\multicolumn{3}{l}{\textbf{\small Species Table}}\\
\hline
Field & Type & Example \\
\hline
\textsc{\makebox[0pt][l]{species-index}\phantom{target-edge-index}} & integer & $i$ \\
\textsc{species-value} & \makebox[0pt][l]{function  (typ.\ positive)}\phantom{real (typically positive)}& $X^i(t)$ \\
\hline
\multicolumn{3}{|l|}{Unique key: \textsc{species-index}}\\
\hline
\end{tabular}

%% file: interaction-table.tex
\begin{tabular}{|l|l|l|}
\multicolumn{3}{l}{\textbf{\small Interaction Table}}\\
\hline
Field & Type & Example \\
\hline
\textsc{\makebox[0pt][l]{interaction-index}\phantom{target-edge-index}} & integer & $a$\\
\textsc{rate-function} & \makebox[0pt][l]{function (typ.\ positive)}\phantom{real (typically positive)} & $f_a(X)$ \\
\hline
\multicolumn{3}{|l|}{Unique key: \textsc{interaction-index}}\\
\hline
\end{tabular}

%% file: target-table.tex
\begin{tabular}{|l|l|l|}
\multicolumn{3}{l}{\textbf{\small Target Table}}\\
\hline
Field & Type & Example \\
\hline
\textsc{interaction-index} & integer & $a$\\
\textsc{target-edge-index} & integer & $m$\\
\textsc{species-index} & integer & $i$\\
\textsc{target-weight} & real (typically positive)& $\alpha_{a,m}$ \\
\hline
\multicolumn{3}{|l|}{Unique key: \{\textsc{interaction-index}, \textsc{target-edge-index}\}}\\
\multicolumn{3}{|l|}{Foreign keys: \textsc{interaction-index}, \textsc{species-index} }\\
\hline
\end{tabular}

%% file: source-table.tex
\begin{tabular}{|l|l|l|}
\multicolumn{3}{l}{\textbf{\small Source Table}}\\
\hline
Field & Type & Example \\
\hline
\textsc{interaction-index} & integer & $a$\\
\textsc{\makebox[0pt][l]{source-edge-index}\phantom{target-edge-index}} & integer & $n$\\
\textsc{species-index} & integer & $i$\\
\textsc{source-weight} & real (typically positive)& $\beta_{a,n}$ \\
\hline
\multicolumn{3}{|l|}{Unique key: \{\textsc{interaction-index}, \textsc{source-edge-index}\}}\\
\multicolumn{3}{|l|}{Foreign keys: \textsc{interaction-index}, \textsc{species-index} }\\
\hline
\end{tabular}